\documentclass[aps,twocolumn,epsfig,groupedaddress,showpacs,floatfix]{revtex4}
                                                                                
\usepackage{color}
                                                                                
\usepackage{graphicx}
\usepackage{epsfig}
                                                                                
\usepackage{amssymb}
\usepackage{amsmath}

\def\menorsim{\smash{\mathop{<}\limits_{\raise3pt\hbox{$\sim$}}}}
\def\maiorsim{\smash{\mathop{>}\limits_{\raise3pt\hbox{$\sim$}}}}

\begin{document}


                                                                                

\title{Net-Baryon Physics: Basic Mechanisms}



\author{J. Alvarez-Mu\~niz}
\affiliation{IGFAE and Dep. Fisica Particulas, Univ. Santiago de Compostela, 
15782 Santiago de Compostela, Spain}
\author{R. Concei\c{c}\~ao}
\author{J. Dias de Deus}
\author{M.C. Esp\'{\i}rito Santo}
\author{J. G. Milhano}
\author{M. Pimenta}
\email[Corresponding author]{, pimenta@lip.pt}
\affiliation{CENTRA, LIP and IST, Av. Rovisco Pais, 1049-001 Lisboa, Portugal}


\date{\today}

\begin{abstract}
It is well known that, in nuclear collisions, a sizable fraction of the available energy 
is carried away by baryons. As the baryon number is conserved, the net-baryon $B-\bar{B}$ 
retains information on the energy-momentum carried by the incoming nuclei.
A simple but consistent model for net-baryon production in high energy hadron-hadron, 
hadron-nucleus and nucleus-nucleus collisions is presented.
The basic ingredients of the model are valence string formation based on standard PDFs 
with QCD evolution and string fragmentation via the Schwinger mechanism.
The results of the model are presented and compared with both data 
and existing models.
These results show that a good 
description of the main features of net-baryon data is possible on the framework of a 
simplistic model, with the advantage of making the fundamental production mechanisms manifest.
\end{abstract}




\maketitle

\section{Introduction}
\label{sec:introd}

In hadron-hadron, hadron-nucleus and nucleus-nucleus interactions a sizable fraction of the available energy 
in a collision is carried away by baryons~\cite{netbar,vanhove,feynman}. 
As the baryon number is conserved, the measured net-baryon, $B-\bar{B}$, 
keeps track of the energy-momentum carried by the incoming particles. 
An important question to be asked is: how does the fraction of energy carried by the net-baryon evolve as a function 
of the centre-of-mass collisional energy per nucleon, $\sqrt{s}$~?
This question is important because, on one hand, a decrease of the fraction of energy going into the net-baryon implies more 
energy available to create the deconfined quark-gluon state of matter, and on the other hand, such a decrease may reduce the 
possibility of producing fast particles in very high energy cosmic ray experiments.
For more than 30 years, since the ISR at CERN, particle production studies have been limited to mid rapidity. Fortunately, 
with RHIC, large rapidity data became available, and, hopefully, the same will happen for the LHC.
In fact, if one does not measure the physics at high rapidity the most elementary physical constraint, namely energy conservation, 
cannot be applied~\cite{GM-JDD,Back}.

In most of the existing Monte Carlo models~\cite{qgsjet1,qgsjet2,epos,sibyll}, 
the physics of net-baryon production is very much obscured by the complexity of extensive 
and detailed codes. Often, the basic production mechanisms do not appear in a transparent way. 
In this paper we present a simple but consistent model for net-baryon production in high energy hadron-hadron (h-h), 
hadron-nucleus (h-A) and nucleus-nucleus (A-A) collisions.
As it happens in most of the existing codes, we shall work in the framework of the Dual Parton Model (DPM)~\cite{capella} 
with string formation (valence strings, in the present case) based on standard Parton Distribution Functions (PDFs) 
with QCD evolution and string fragmentation via the Schwinger mechanism.
The basic ingredients of the model are:
\begin{itemize}
\item Formation of extended color fields or strings, making use of PDFs for valence quarks;
\item Evolution with momentum transfer $Q^2$, as a consequence of the QCD evolution of the PDF's;
\item Fragmentation of each string with formation of a fast baryon (net-baryon) and other particles.
\end{itemize}
This paper is organized as follows. In section~\ref{sec:current}, the predictions of current models and the available experimental 
data are briefly reviewed and compared. In section~\ref{sec:themodel}, our simple net-baryon model is described. The results of 
the model are presented and compared with data and with existing models in section~\ref{sec:results}. 
A summary and brief conclusions are presented in section~\ref{sec:conclusions}.

\section{Current model predictions and experimental data}
\label{sec:current}

\begin{figure}[htb]
\begin{center}
\fontsize{9pt}{9pt}
\epsfig{figure=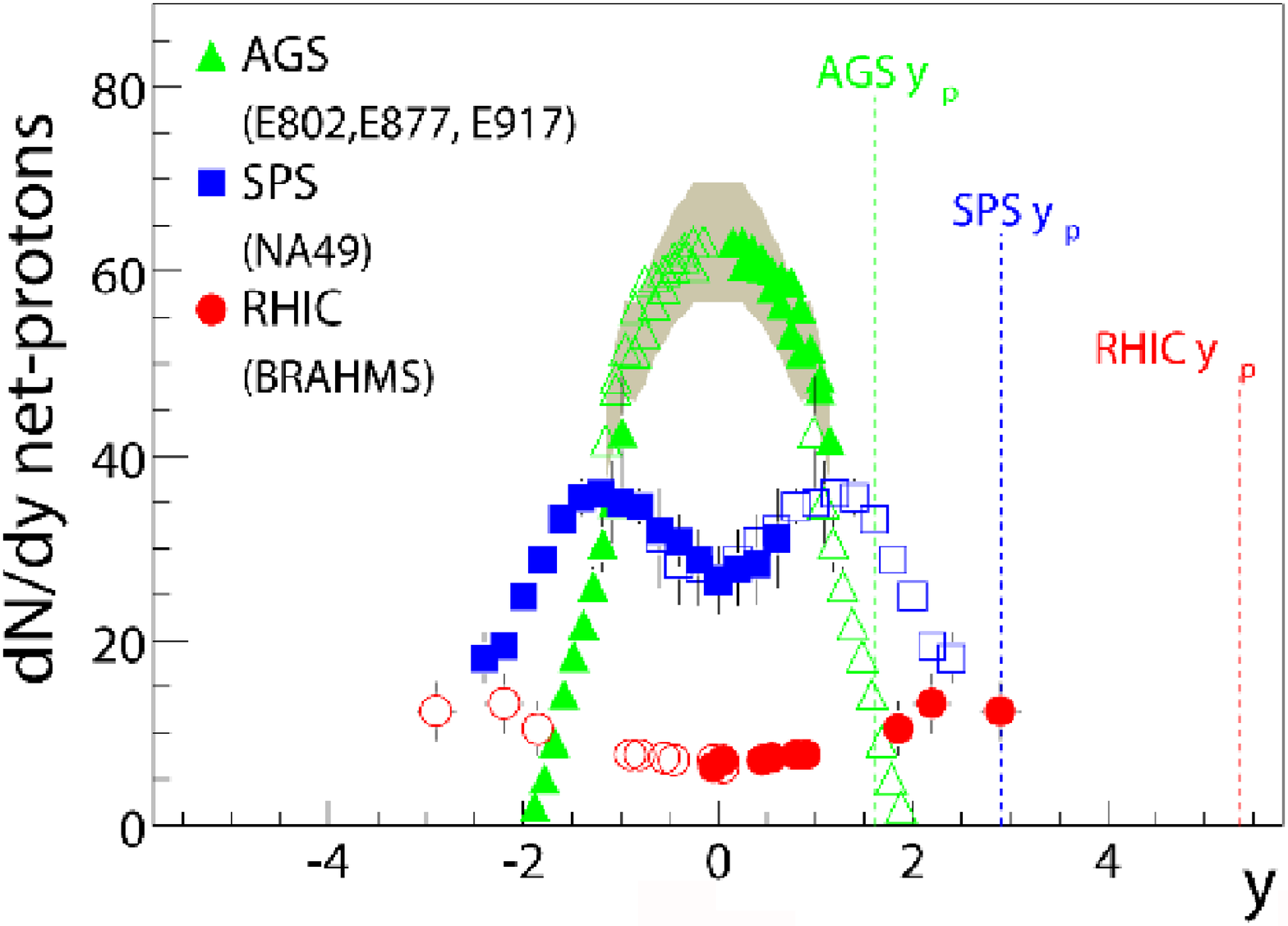, angle=0, width=0.4\textwidth}
\epsfig{figure=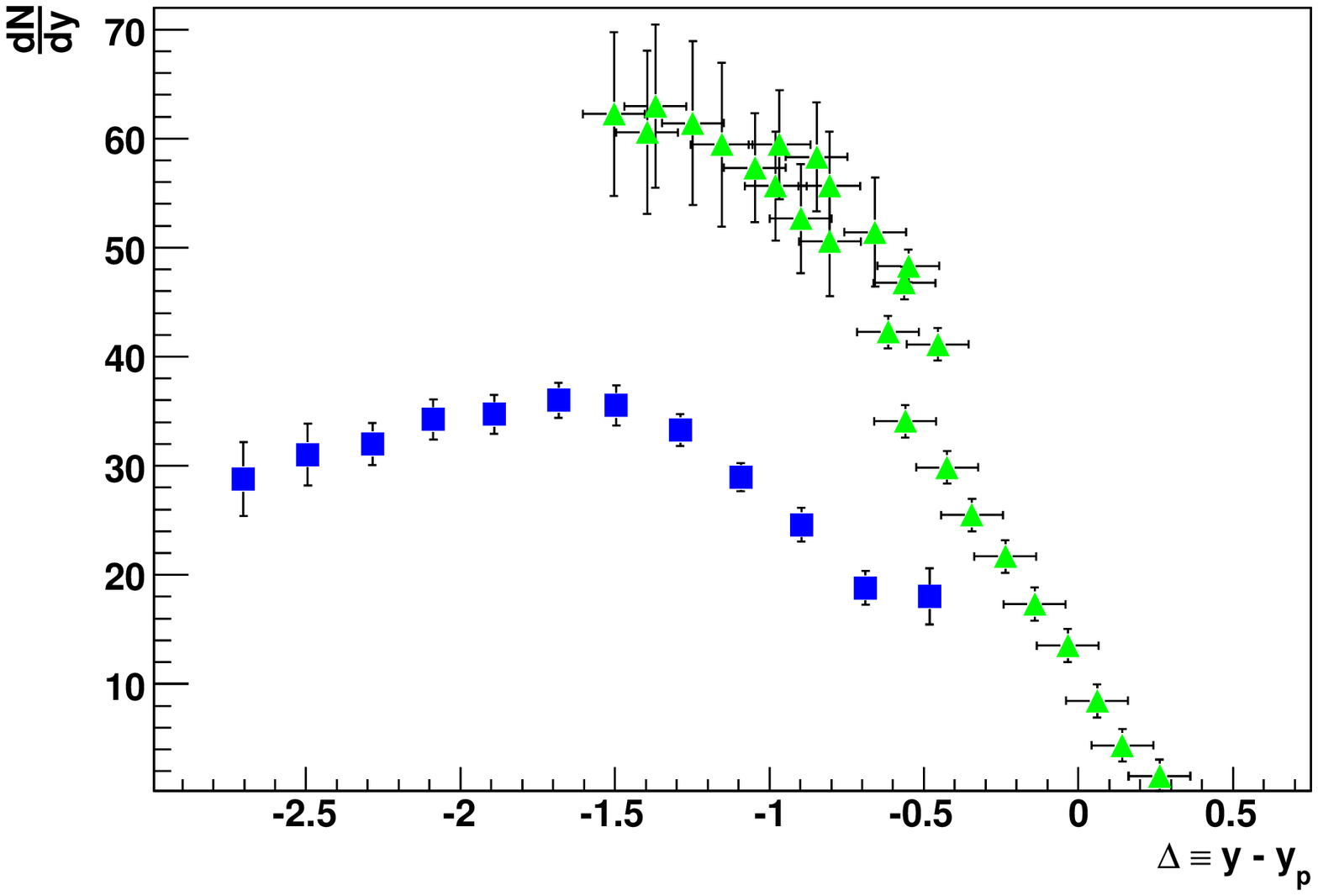, angle=0, width=0.4\textwidth}
\caption{The net-proton distribution at AGS~\cite{ags} (Au+Au at $\sqrt{s} \simeq 5$~GeV), 
SPS~\cite{sps} (Pb+Pb at $\sqrt{s} \simeq 17$~GeV), and RHIC (Au+Au at $\sqrt{s}=200$~GeV) is shown (upper plot). 
The data correspond to the 5\% most central collisions and the errors are both statistical and systematic 
(the light gray band shows the 10\% overall normalization uncertainty on the E802 points, but not the 15\% for E917). The data have been 
symmetrised. For RHIC data, filled points are measured and open points are symmetrised, while the opposite is true for AGS and SPS data 
(for clarity). At AGS weak decay corrections are negligible and at SPS they have been applied. Taken from~\cite{rhic-brahms}.
The distribution of $\Delta \equiv y - y_p$ is shown (lower plot) for SPS and AGS data.}
\label{netbar_art}
\end{center}
\end{figure}

\begin{figure}[htb]
\begin{center}
\fontsize{9pt}{9pt}
\epsfig{figure=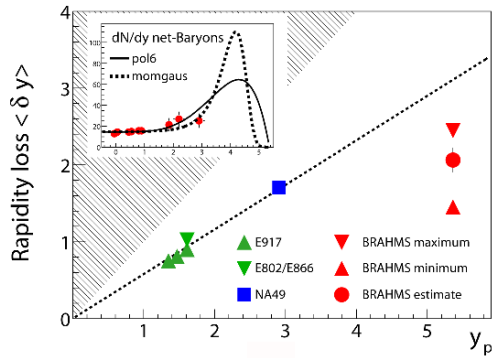, angle=0, width=0.4\textwidth}
\caption{Rapidity loss as a function of beam rapidity (in the CM). The hatched area indicates the unphysical region, 
and the dashed line shows the phenomenological scaling $\left< \delta y \right> = 0.58 y_p$. 
The inset plot shows the BRAHMS net-baryon distribution (data points) with fits (lines) needed to extrapolate up to the
beam rapidity, explaining the large uncertainty associated to the BRAHMS rapidity loss measurement. 
Taken from~\cite{rhic-brahms}.}
\label{raploss_art}
\end{center}
\end{figure}

The data presently available on net-proton or net-baryon production is scarce. 
The most recent results are from RHIC~\cite{rhic-brahms}. 
These results are presented both in terms of net-baryon rapidity $y$ and in terms of rapidity loss, 
defined as
\begin{equation}
\left< \delta y  \right> = y_p - \left< y \right>\, ,
\end{equation}
where $y_p$ is the beam rapidity and $\left< y \right>$ is the mean net-baryon rapidity after the collision, 
given by
\begin{center}
\begin{equation}
\left< y \right> = \frac{2}{N_{part}} \int _{0}^{y_p} y \frac{dN_{B-\bar{B}}(y)}{dy} dy\, .
\label{mean_yb}
\end{equation}
\end{center}
Here, $N_{part}$ is the number of participants in the collision and $N_{B-\bar{B}}$ is the net-baryon number.
It is worth noting that the net-baryon results depend directly on the number of participants in the collision
and thus on the collision centrality. 
The relation between the number of participants and the impact parameter of the collision can be established, 
for instance, in the context of the Glauber model~\cite{glauber}. 

The net-baryon rapidity distributions at different centre-of-mass energies for different beam-target systems were 
compared in~\cite{rhic-brahms}.
In figure~\ref{netbar_art},
the net-proton distributions at AGS (Au+Au at $\sqrt{s} \simeq 5$~GeV)~\cite{ags}, 
SPS (Pb+Pb at $\sqrt{s} \simeq 17$~GeV)~\cite{sps}, 
and RHIC (Au+Au at $\sqrt{s}=200$~GeV)~\cite{rhic-brahms} for central collisions (the 5\% most central) are 
shown (upper plot). 
The distributions show a strong energy dependence, with the mid rapidity region corresponding to a peak at AGS, a dip
at the SPS, and a broad minimum at RHIC. 
The distribution of $\Delta = y - y_p$ is shown (lower plot) for SPS and AGS data.
It is apparent that neither Feynman scaling nor limiting fragmentation are
satisfied in net-baryon production.

In figure~\ref{raploss_art}, the rapidity loss data is shown as a function of the beam rapidity (in the centre-of-mass
frame) for AGS, SPS and RHIC. The large uncertainty associated with the BRAHMS data point is,
as illustrated in the inset plot, related to a relatively unconstrained extrapolation to the high rapidity region,
required for the rapidity loss calculation.   

\begin{figure}[htb]
\begin{center}
\fontsize{9pt}{9pt}
\epsfig{figure=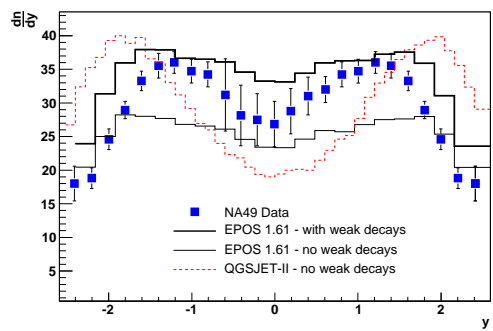, angle=0, width=0.4\textwidth}
\caption{The net-proton rapidity distributions from EPOS 1.61 (full lines) and QGSJET-II.03 (dashed line) for 
Pb-Pb collisions at $\sqrt{s} \simeq 17$~GeV are shown and compared with data NA49 SPS data (points).
For EPOS, the results obtained including (thick line) and excluding (thin line) the
strangeness contribution are shown. For QGSJET, the ``no-decays'' curve is shown.
For details on the data see figure~\ref{netbar_art}}.
\label{netbar_mod}
\end{center}
\end{figure}

Let us now turn to net-baryon production as implemented in the existing Monte Carlo models.
QGSJET-II~\cite{qgsjet2} and EPOS~\cite{epos} are amongst the presently most widely used hadronic 
models in high energy and cosmic ray physics. 
To our knowledge, there is no systematic study comparing the predictions of these two models on 
net-baryon production between themselves or with experimental data. 

In figure~\ref{netbar_mod}, net-proton rapidity distributions obtained with QGSJET-II.03 and EPOS 1.61
for Pb-Pb collisions at $\sqrt{s} \simeq 17$~GeV are shown and compared with experimental data. 
According to~\cite{sps} weak decay corrections have been applied to the data.
Following~\cite{Pb-centrality}, the impact parameter range (0 to 3.1 fm) corresponding to the centrality
cut of 5\% applied to data was selected in QGSJET and EPOS. 
For EPOS, the results obtained leaving all particles free to decay and contribute 
to the net-baryon are shown together with those obtained switching off all weak decays.
The ``all-decays'' EPOS curve reproduces the trend seen in data, with an excess that could
be due to the presence of net-baryon from weak decays. However, the strangeness
effect compensation in the ``no-decays'' curve seems to be too strong.
For QGSJET-II, the ``no-decays'' curve is shown (and was found to be very similar to what is 
obtained from the model by default).
One should recall that QGSJET-II is not expected to perform very well at such low energies.


\section{The model}
\label{sec:themodel}

\begin{figure}[htb]
\begin{center}
\fontsize{9pt}{9pt}
\epsfig{figure=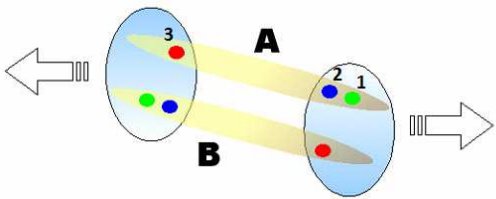, angle=0, width=0.4\textwidth}
\caption{Schematic representation of a proton-proton collision, with the formation of two valence strings.}
\label{collision}
\end{center}
\end{figure}

Let us now describe a simple model for net-baryon production. In the spirit of~\cite{vanhove,feynman,capella,bass},
the basic assumption is that net-baryon production in proton-proton collisions is strongly correlated with 
the formation and fragmentation of two color singlet strings, each one with two valence quarks from one of the protons, 
and one valence quark from the other proton. This is schematically shown in figure~\ref{collision}. 

\begin{figure}[htb]
\fontsize{9pt}{9pt}
\begin{center}
\begin{center}
\epsfig{figure=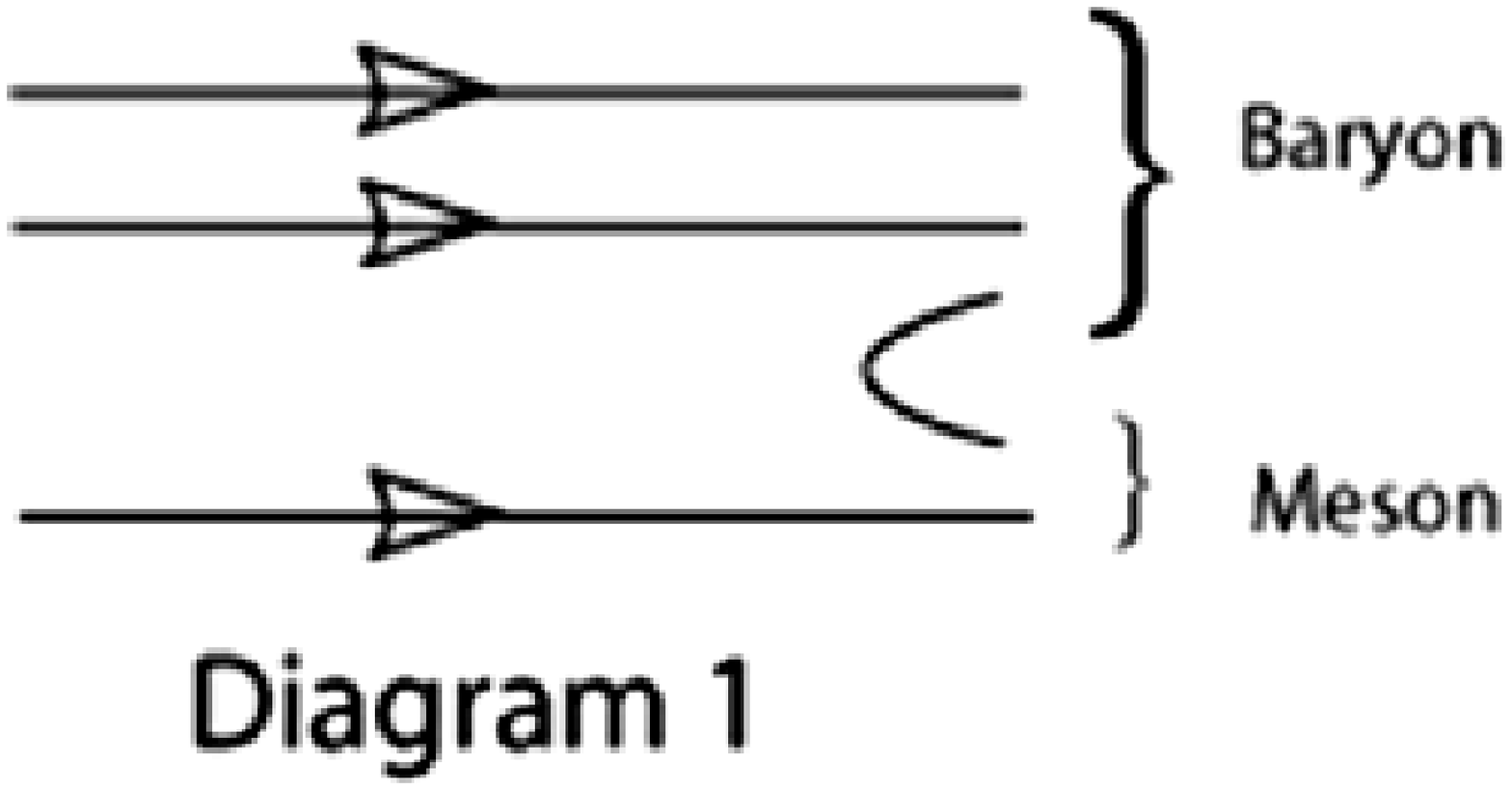, angle=0, width=0.3\textwidth}
\scalebox{1.2}{\bf $x_1 - x_2 < x_2 - (-x_3)$}
\end{center}
\vspace{0.5 cm}
\begin{center}
\epsfig{figure=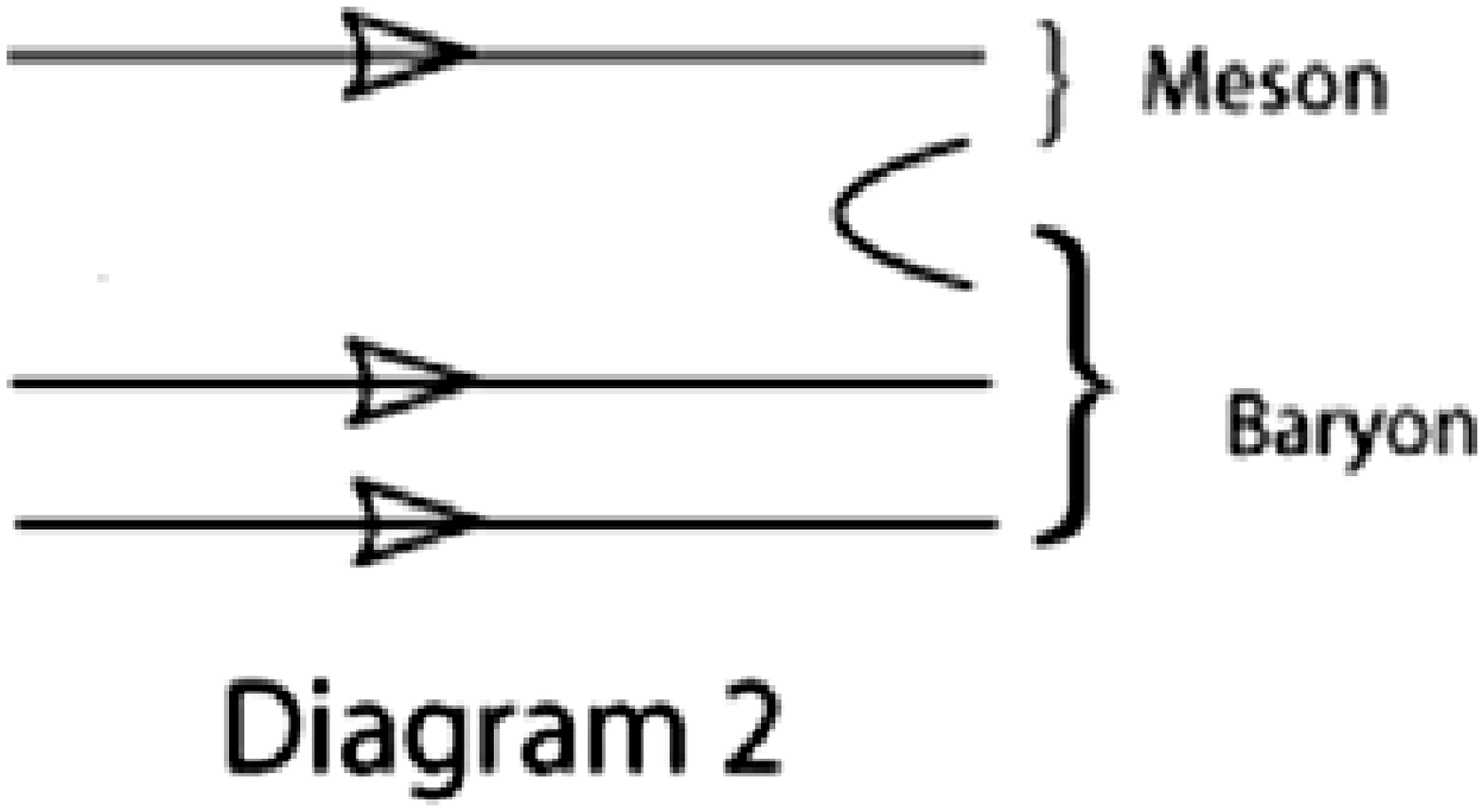, angle=0, width=0.3\textwidth}
\scalebox{1.2}{\bf $x_1 - x_2 < x_2 - (-x_3)$}
\end{center}
\caption{The two main valence string fragmentation diagrams.} 
\label{dpmff}
\end{center}
\end{figure}

Referring to the figure,
let $x_1$, $x_2$ and $x_3$ be the fractions of momentum carried by the valence quarks forming string A. 
Quarks 1 and 2 are from the proton with positive momentum in the proton-proton reference frame, 
and quark 3 is from the other proton. No transverse momentum is considered within this model.
By choice, $x_1 > x_2 > x_3$, with $x_3<0$. 
The energy and momentum of each string are obtained adding directly the energy and momentum 
carried by each of the valence quarks. For string A
\begin{center}
\begin{eqnarray}
E_{string} &=& (x_1 + x_2 + (-x_3))\frac{\sqrt{s}}{2}\, ,\\
P_{string} &=& (x_1 + x_2 - (-x_3))\frac{\sqrt{s}}{2}\, ,\\
M_{string} &=& \sqrt{(x_1 + x_2)\ (-x_3)\ s}\, ,
\end{eqnarray}
\end{center}

where the momentum fractions $x_1$, $x_2$ and $x_3$ are determined using the quark PDFs.
For each $\sqrt{s}$, an effective momentum transfer $Q^2$ adjusted from experimental data was chosen
(see section~\ref{sec:results} for details).
In this work the CTEQ6M parton distribution functions~\cite{cteq6} were used.

The simplest possible model for fragmentation is assumed. Each string decays into a baryon and a meson in the following way:
the string is cut in two pieces and a $q \bar{q}$ pair is formed, from the vacuum, either between quarks 2 and 3 (belonging
the different protons) or between quarks 1 and 2 (belonging to the same proton, with positive momentum in this string A example).
The quark pair with the largest momentum difference is chosen. 
The string piece that inherits two valence quarks originates the baryon, whereas the string piece that inherits one valence 
quark originates the meson. 
This mechanism corresponds to the diagrams represented in figure~\ref{dpmff}.
The weights of the two diagrams are, 
in this simple model, given only by kinematics. 
For string A the first diagram will be more probable (especially for low $\sqrt{s}$). However, the weight of the second diagram 
can be as large as $40 \%$ above LHC energies.

The $q \bar{q}$ pair formed from the vacuum was taken to be either a $u \bar{u}$ or a $d \bar{d}$,
and the full quark combinatorics was then performed in order to determine the nature of the possible 
outcoming baryon. Both fundamental and excited states were considered, taking spin-dependent weights (2j+1). 
The decays of the unstable baryons were then performed and the outcoming nucleons included in the 
net-baryon calculations.
The contribution from $s$ quarks was not considered here. It was estimated from 
the Schwinger model to be about 25\%.

\begin{figure}[htb]
\begin{center}
\fontsize{9pt}{9pt}
\epsfig{figure=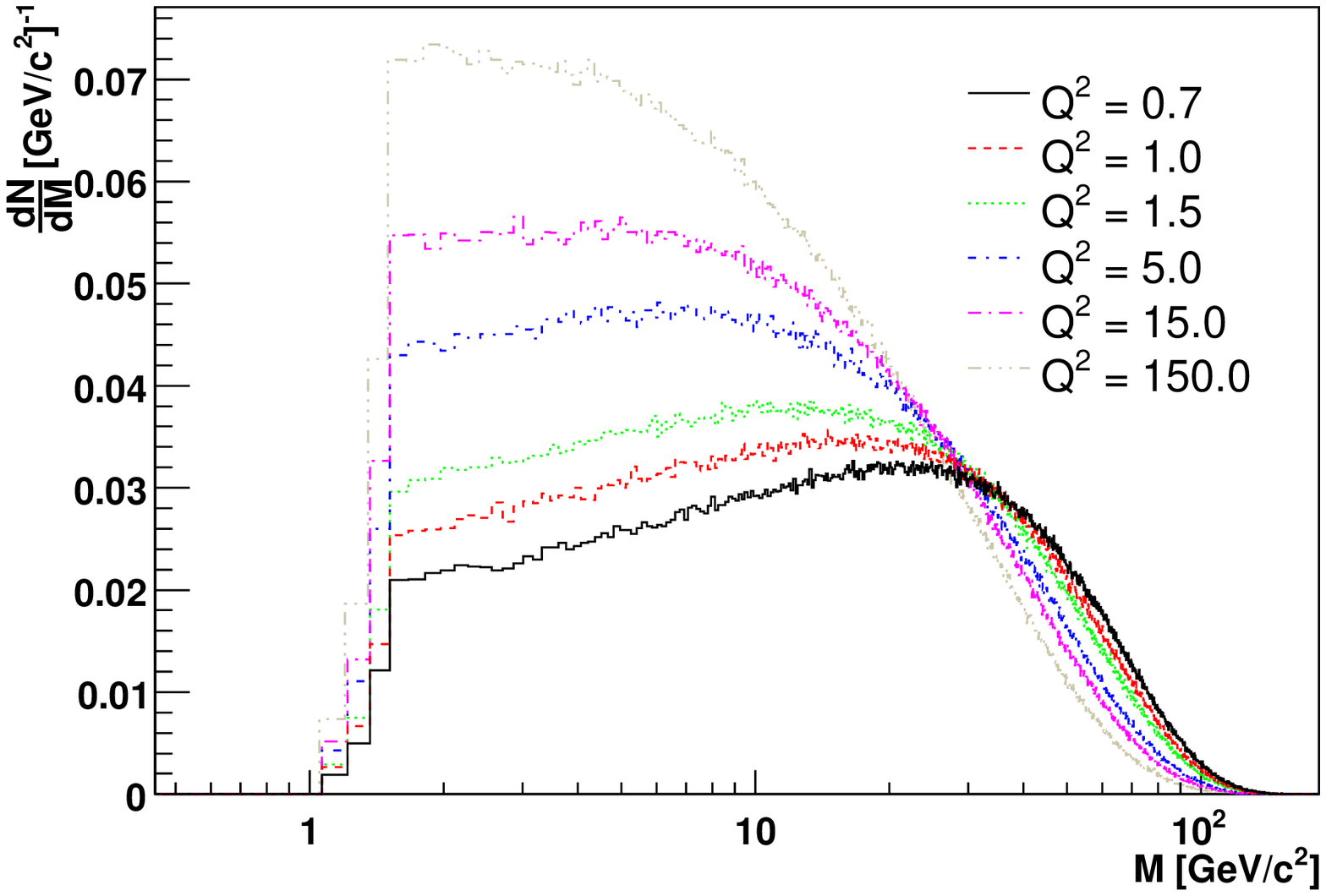, angle=0, width=0.4\textwidth}
\epsfig{figure=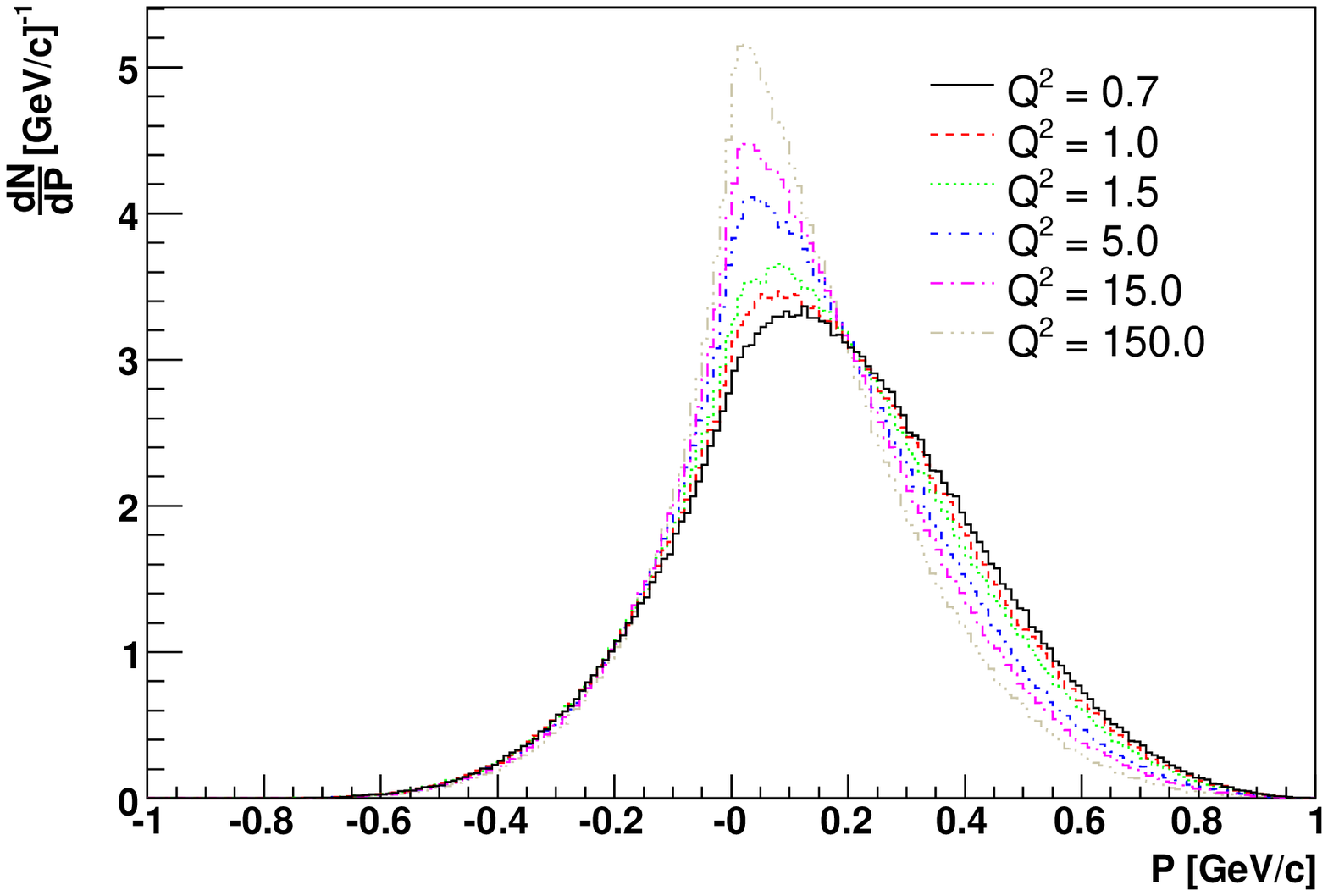, angle=0, width=0.4\textwidth}
\caption{Evolution with $Q^2$ of the string mass (upper plot) and momentum (lower plot)
distributions for $\sqrt{s}=200$~GeV.
The mass distribution applies to both valence strings, while the momentum distribution for the case of 
string B is simply symmetric.}
\label{string200}
\end{center}
\end{figure}

\begin{figure}[htb]
\begin{center}
\fontsize{9pt}{9pt}
\epsfig{figure=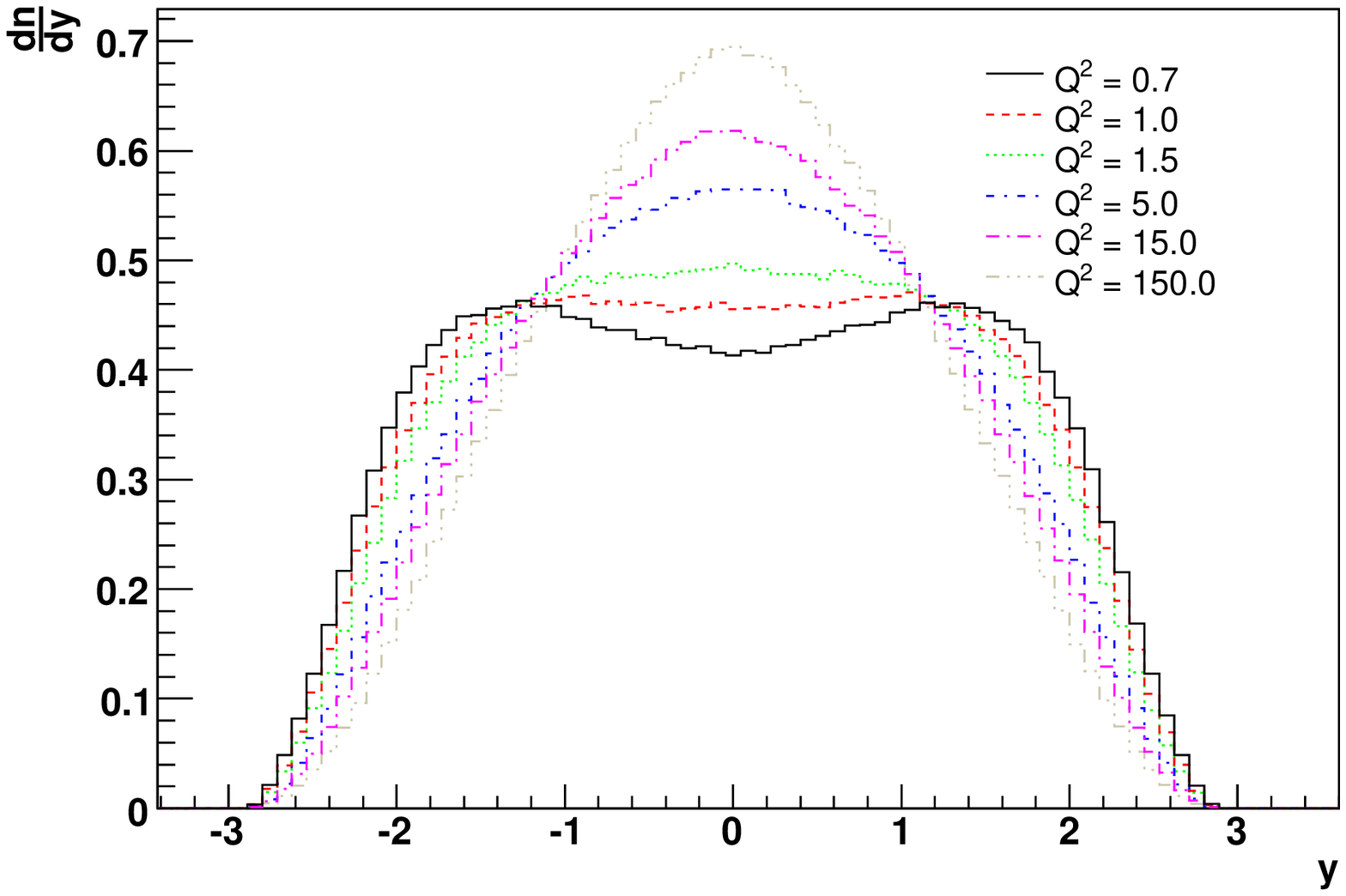, angle=0, width=0.4\textwidth}
\epsfig{figure=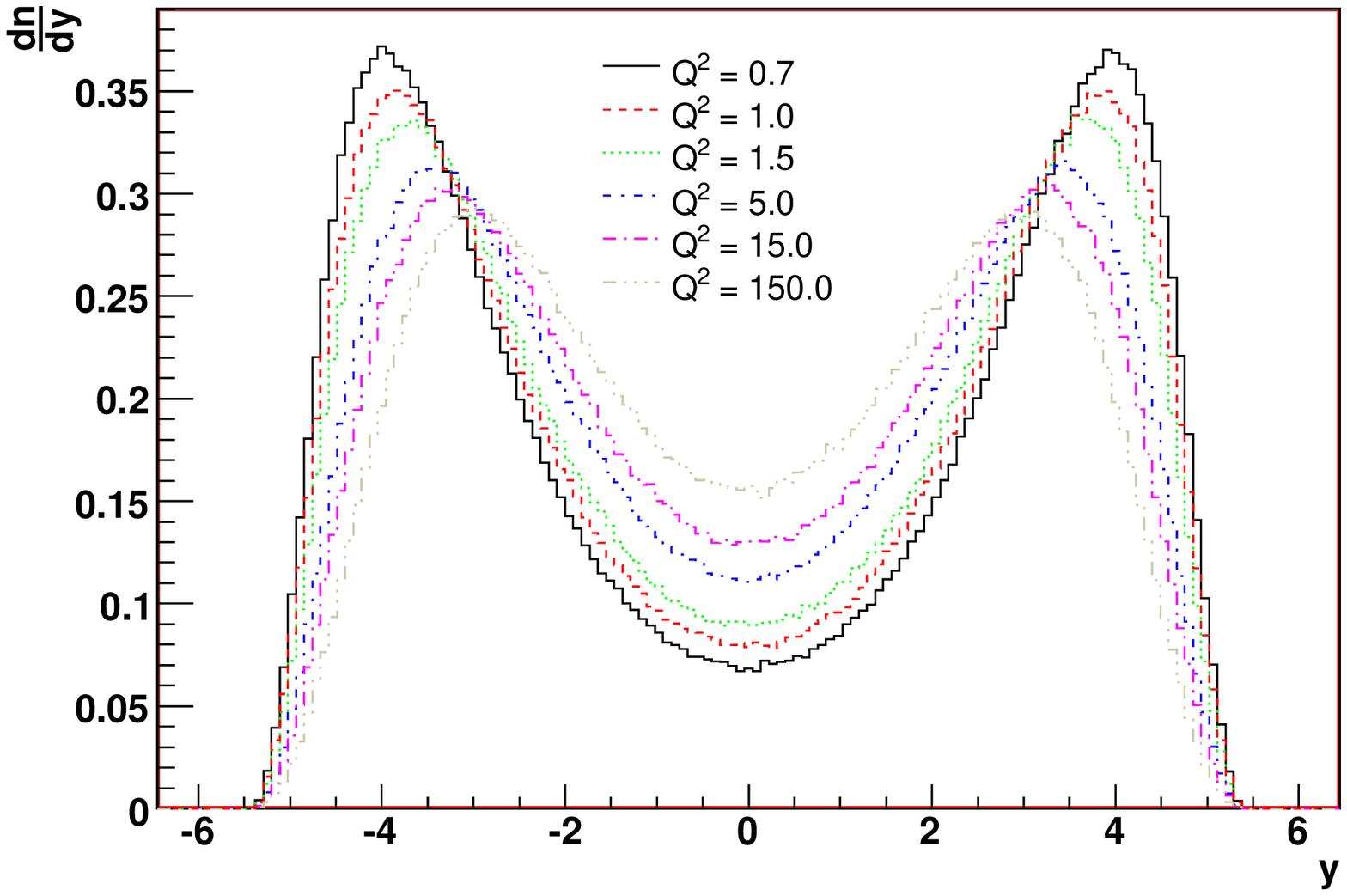, angle=0, width=0.4\textwidth}
\caption{Evolution of the net-baryon rapidity with $Q^2$ at $\sqrt{s}=17$~GeV (upper plot) and 
$\sqrt{s}=200$~GeV (lower plot).}
\label{netbar}
\end{center}
\end{figure}

The string mass and momentum distributions as function of $Q^2$ are given in figure~\ref{string200}. 
The mass distribution applies to both valence strings, while the momentum distribution for the case of 
string B is simply symmetric.
The net-baryon rapidity distributions for different values of $Q^2$
are presented in figure~\ref{netbar}, for two different centre-of-mass energies. 
This figure illustrates not only the evolution with $Q^2$ but also
the clear effect of kinematics, due to the increase of $\sqrt{s}$, 
on the main features of the distribution.

It is worth noting that the inclusion of diagram 2, in addition to diagram 1, with weights determined
simply by kinematics, reproduces some of the effects predicted in models with string 
junctions~\cite{junctions,shabelski} or popcorn~\cite{popcorn} mechanisms for the transport of baryon number 
from the beam rapidity into the central region $y \sim 0$. 
These effects can thus be achieved in a simple DPM model, based
on valence strings and a $Q^2$ parameterisation.

For the purpose of estimating the net-baryon production in A-A collisions, 
we shall use the simple approximation that $dn/dy(B-\bar{B})$ at a given $\sqrt{s}$ is proportional
to the number of participating nucleons, $N_{part}$,

\begin{equation}
\label{eq:AA}
\frac{dn}{dy} \Bigr \rvert_{A-A} \simeq \frac{1}{2}N_{part} \times \frac{dn}{dy} \Bigr \rvert _{p-p}.
\end{equation}

This relation is presumably reasonable as net-baryon production is originated from valence strings
associated to wounded (or participating) nucleons. 
At mid rapidity, saturation effects originate additional production of mesons and baryons. This
affects $(B + \bar{B})$, but not $B - \bar{B}$ production.
An attempt to estimate
nuclear effects correction factors for the valence quark PDFs was made using EKS98~\cite{eks98} and 
nDS~\cite{nds}. Values below 10-15\% were found.

\section{Results}
\label{sec:results}

In order to fully define the model, we need to choose an effective $Q^2$ value as input for the quark PDFs.
This is done by adjusting the results of the model to the experimental data, i.e., by
choosing, for each $\sqrt{s}$, the effective $Q^2$ for which the model better describes the data. 
In the fitting procedure, the global normalisation factor (the number of participants, 
relying on the approach of eq.~(\ref{eq:AA}) is left as a second free parameter.
In order to compare the results of this model on net-baryon to the experimental data
on net-proton, we consider that net-proton is roughly 1/2 of $(B - \bar{B})$.

\begin{figure}[htb]
\begin{center}
\fontsize{9pt}{9pt}
\epsfig{figure=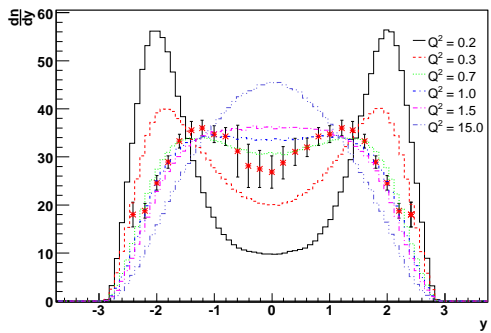, angle=0, width=0.4\textwidth}
\caption{Net-baryon predictions of the present model for different values of $Q^2$, 
and comparison with data, at $\sqrt{s} \simeq 17$~GeV.}
\label{adjust}
\end{center}
\end{figure}

Figure~\ref{adjust} illustrates, for a specific $\sqrt{s}$ and fixing the number of participants, the 
ability of the model to generate very different net-proton rapidity distributions, 
simply by choosing different values of the effective $Q^2$.
At $\sqrt{s}=200$~GeV and $\sqrt{s} \simeq 17$~GeV all the data points (see figure~\ref{netbar_art}) 
were included in the fit. At  $\sqrt{s} \simeq 5$~GeV, and due to energy conservation constraints
imposed in the model, only the points up to the nominal beam rapidity were considered.

The values obtained for the effective $Q^2$ and the number of participants $N_{part}$
at the different centre-of-mass energies are given in table~\ref{tab:results}. 

\begin{center}
\begin{table}
\begin{center}
\begin{tabular}{cccc}
\hline
$\sqrt{s}$ (GeV) & Collision & $Q^2$ & $N_{part}$ \\  
\hline
5         & Au-Au   &    $0.36^{+0.10}_{-0.01}$     &   $260\pm20$     \\ 
17        & Pb-Pb   &    $0.64^{+0.17}_{-0.17}$     &   $300\pm20$     \\ 
200       & Au-Au   &    $1.69^{+0.52}_{-0.78}$     &   $280\pm20$     \\ 
\hline
\end{tabular}
\caption{Results of the fit to the effective $Q^2$ and the number of participants
at the different centre-of-mass energies.}
\label{tab:results}
\end{center}
\end{table}
\end{center}
It is worth noting that the fitted $N_{part}$ values are roughly compatible with what is 
expected from the literature, although about 20\% lower. In fact, the average number of participants for 
Pb-Pb collisions at SPS energies is given in~\cite{Pb-centrality} as 362.
Concerning Au-Au collisions, the value is expected to be only slightly lower (estimated as 
344 to 357 at RHIC energies~\cite{rhic-brahms,Au-centrality,Au-centrality2}) and expected
to depend weakly on $\sqrt{s}$, in fair agreement with the  obtained results.
As pointed out in section~\ref{sec:themodel}, in this simple model strangeness contribution to the net-baryon
was not considered, and it amounts to about 25\%, accounting for the values obtained 
for the number of participants. 

\begin{figure}[htb]
\begin{center}
\fontsize{9pt}{9pt}
\epsfig{figure=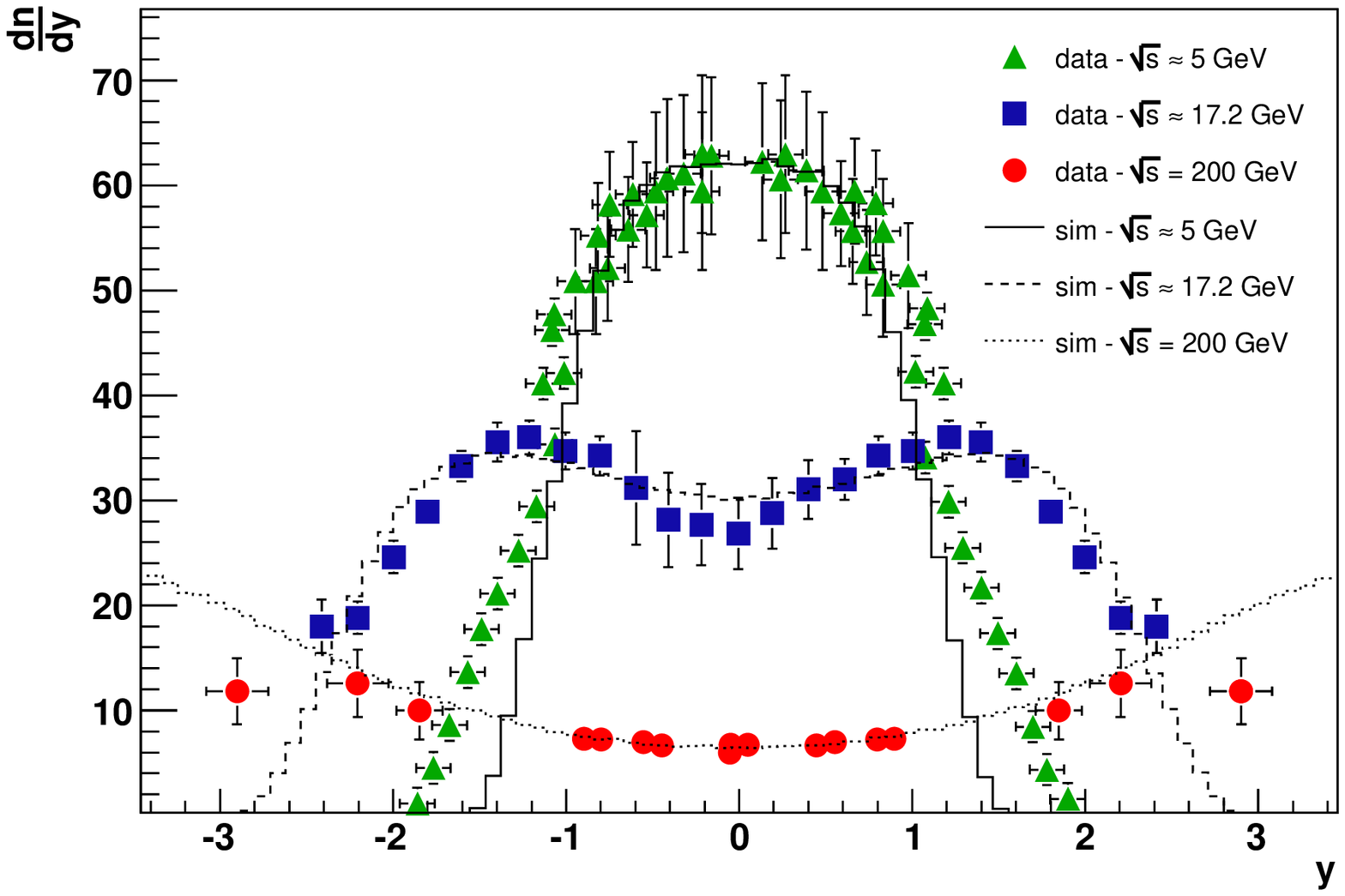, angle=0, width=0.4\textwidth}
\epsfig{figure=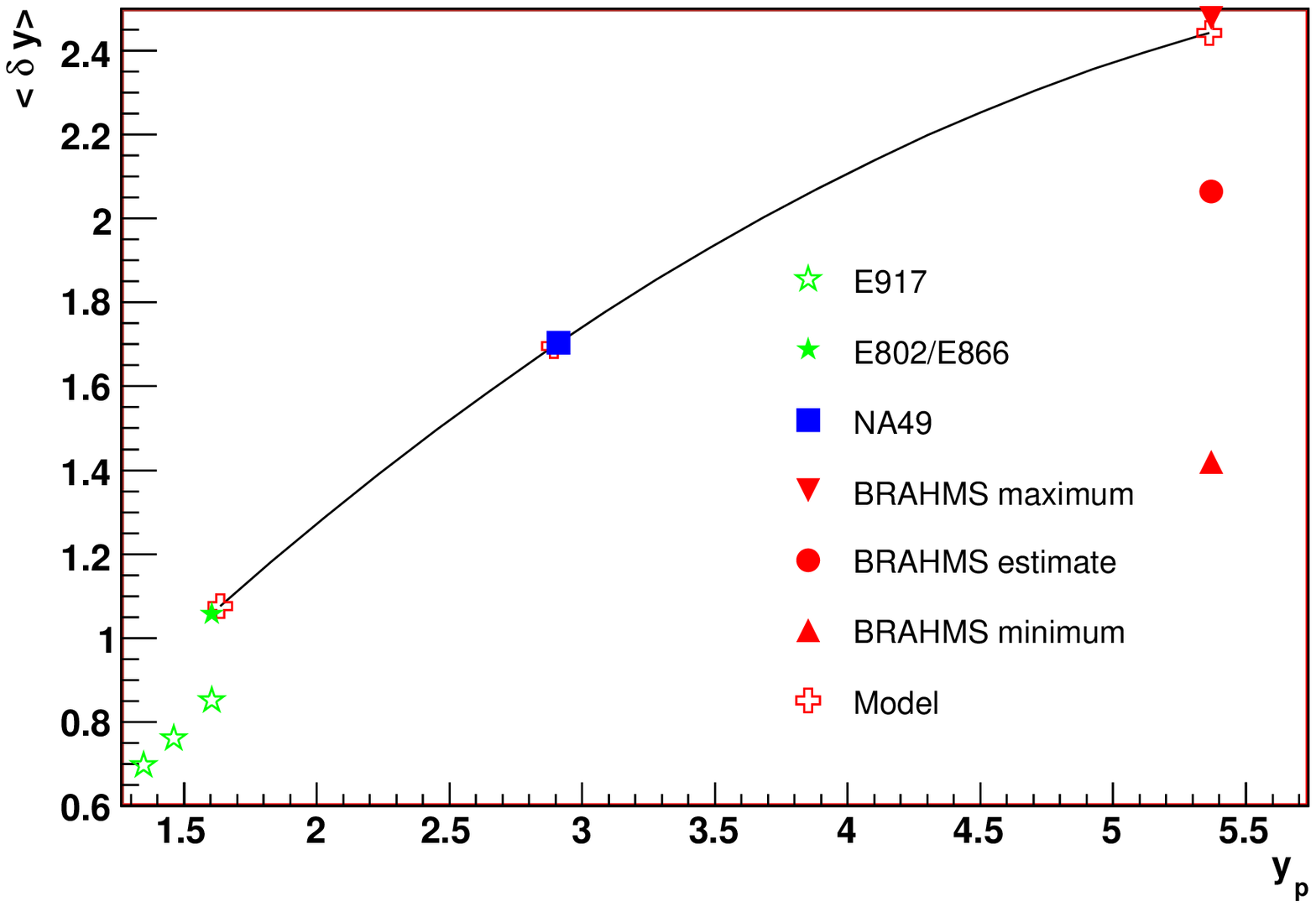, angle=0, width=0.4\textwidth}
\caption{The results of the present model for net-baryon rapidity (upper plot) and
rapidity loss (lower plot) are compared to experimental data at different
centre-of-mass energies. See figure~\ref{netbar_art} for details on the data points.}
\label{results}
\end{center}
\end{figure}

Figure~\ref{results} shows the results of the present model for net-baryon rapidity 
distribution and
rapidity loss, in comparison with experimental data at different centre-of-mass energies.
A fair agreement is found at the measured energies.
For the lowest $\sqrt{s}$, the points above the nominal beam rapidity were 
not considered in the fit.

We can now try to find a relation between the effective $Q^2$ and $\sqrt{s}$.
The  effective $Q^2$ corresponds to the typical transverse size (area) 
of the parton (here, the valence quark). It is reasonable to assume, as in
Regge phenomenology~\cite{regge}, that the average number of partons in a nucleon
increases as a power of the centre of mass energy $\sqrt{s}$. Thus, $\sqrt{s}/Q^2 \sim
R_h^2$, where $R_h$ is the nucleon radius which we take as fixed. It then follows that
$Q^2$ should grow according to

\begin{center}
\begin{equation}
 Q^2 = Q_0^2 \left( \frac{\sqrt{s}}{\sqrt{s_0}} \right)^{\lambda_v} \left[ GeV^2 \right].
\label{eq:lambda}
\end{equation}
\end{center}

The exponent $\lambda_v$ was determined by fitting eq.~(\ref{eq:lambda}) to the available data points
in table~\ref{tab:results}.

\begin{figure}[htb]
\begin{center}
\fontsize{9pt}{9pt} 
\epsfig{figure=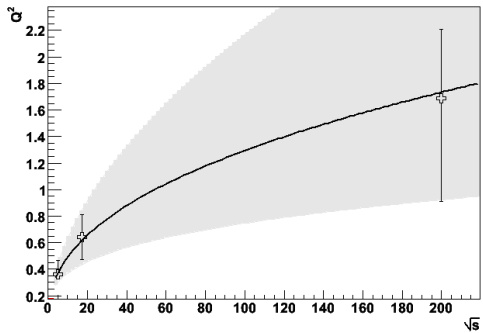, angle=0, width=0.4\textwidth}
\caption{The effective $Q^2$ values at different centre-of-mass energies chosen by tuning the model to
the experimental data are shown. The line shows the fit to the points using eq.~(\ref{eq:lambda})
and the shaded areas correspond to 1$\sigma$ variations of the fit parameters.}
\label{fit_lambda}
\end{center}
\end{figure}

The results are shown in figure~\ref{fit_lambda}, where the data points are the $Q^2$
values adjusted above, the line is the fit with eq.~(\ref{eq:lambda}) and the shaded areas
correspond to 1$\sigma$ variations of the fit parameters.
The obtained results are $\lambda_v=0.42 ^{+0.065}_{-0.120}$ and $Q_0^2 = 0.36 ^{+0.070}_{-0.015}$, taking 
$\sqrt{s_0} = 5$~GeV. It should however be noted that we have just three points
in the fit and the 1$\sigma$ band is rather wide.

\begin{figure}[htb]
\begin{center}
\fontsize{9pt}{9pt}
\epsfig{figure=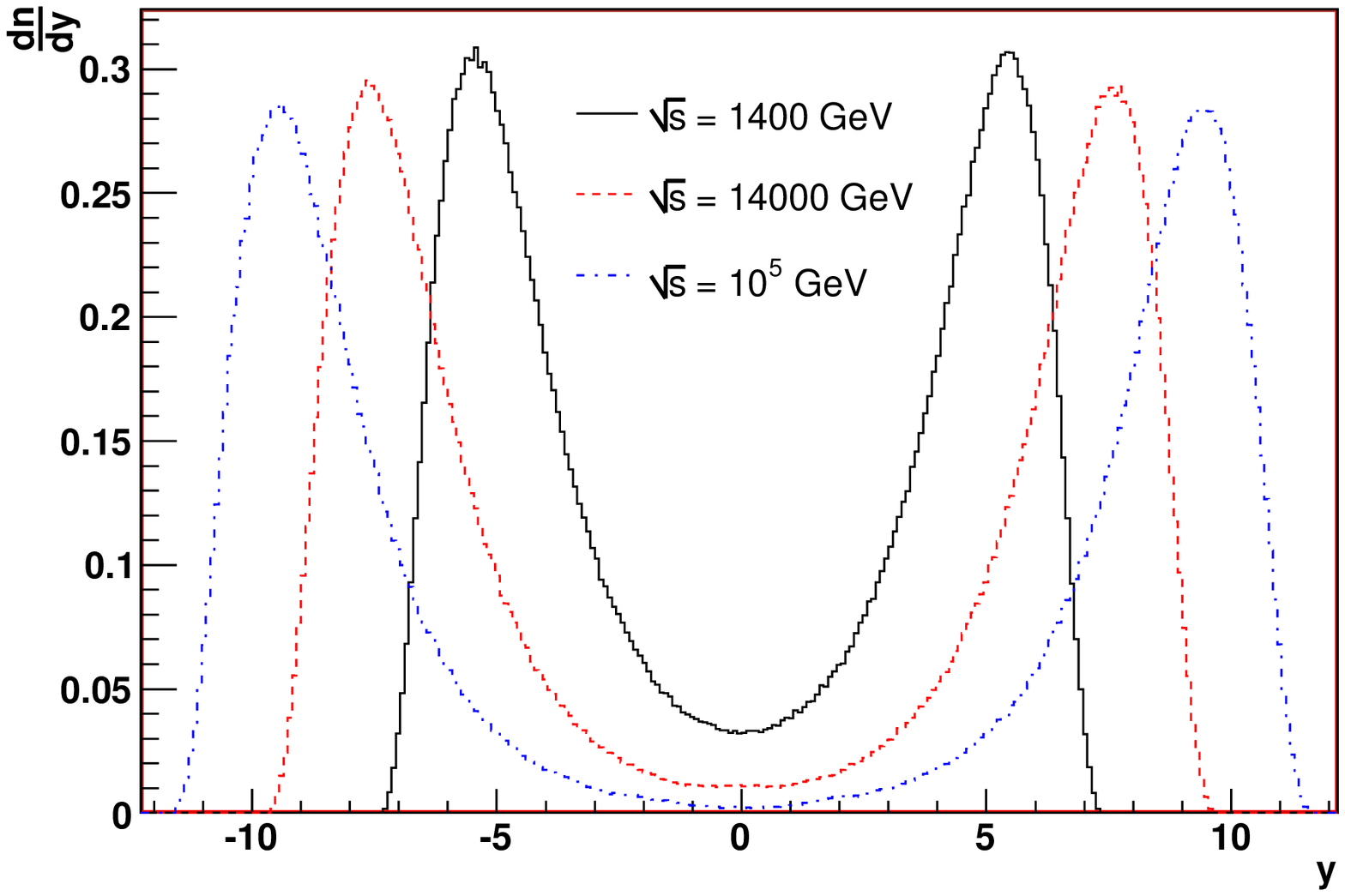, angle=0, width=0.4\textwidth}
\epsfig{figure=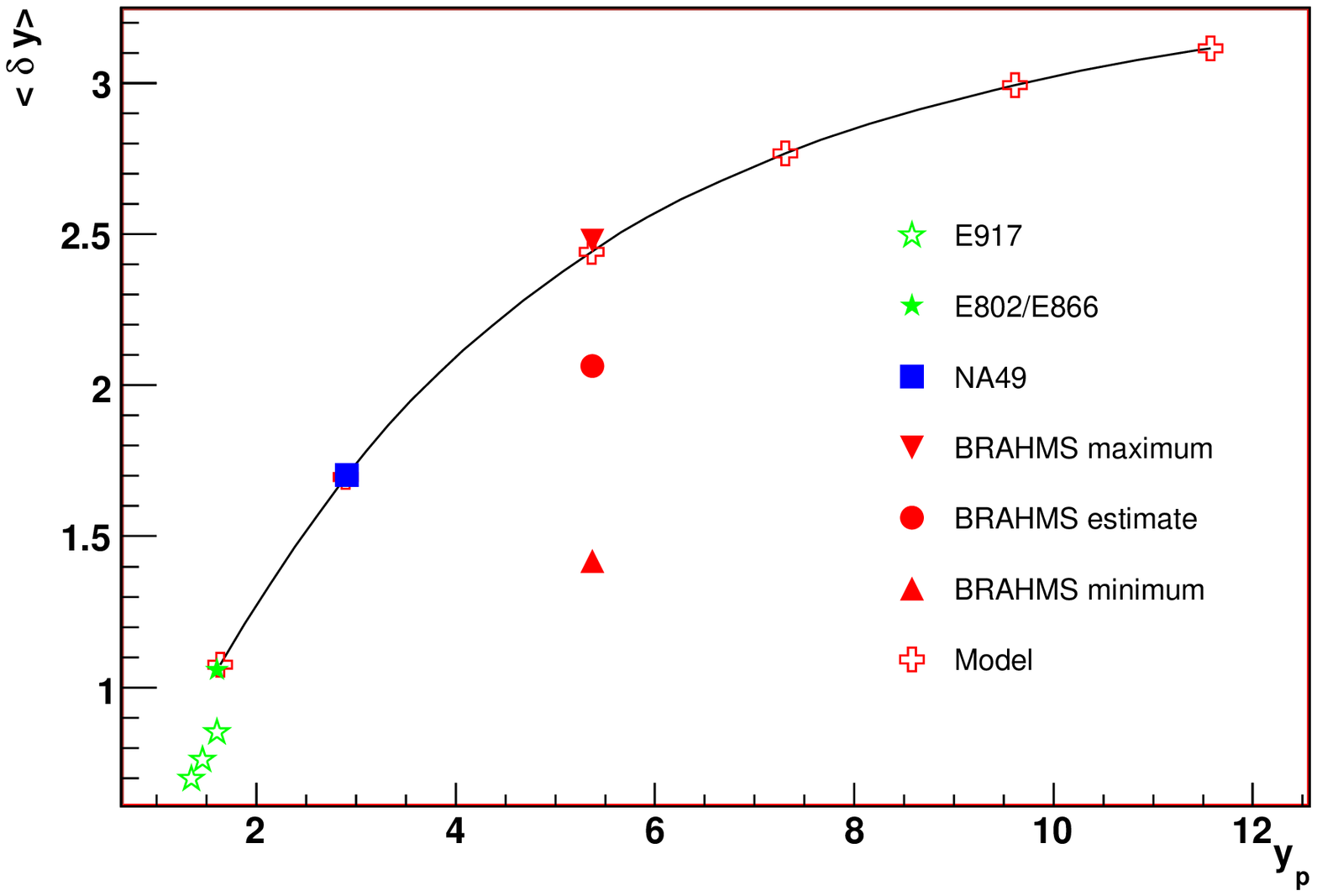, angle=0, width=0.4\textwidth}
\caption{Predictions of the present model for net-baryon rapidity (upper plot) and
rapidity loss (lower plot) including higher energies.}
\label{predictions}
\end{center}
\end{figure}

\begin{figure}[htb]
\begin{center}
\fontsize{9pt}{9pt}
\epsfig{figure=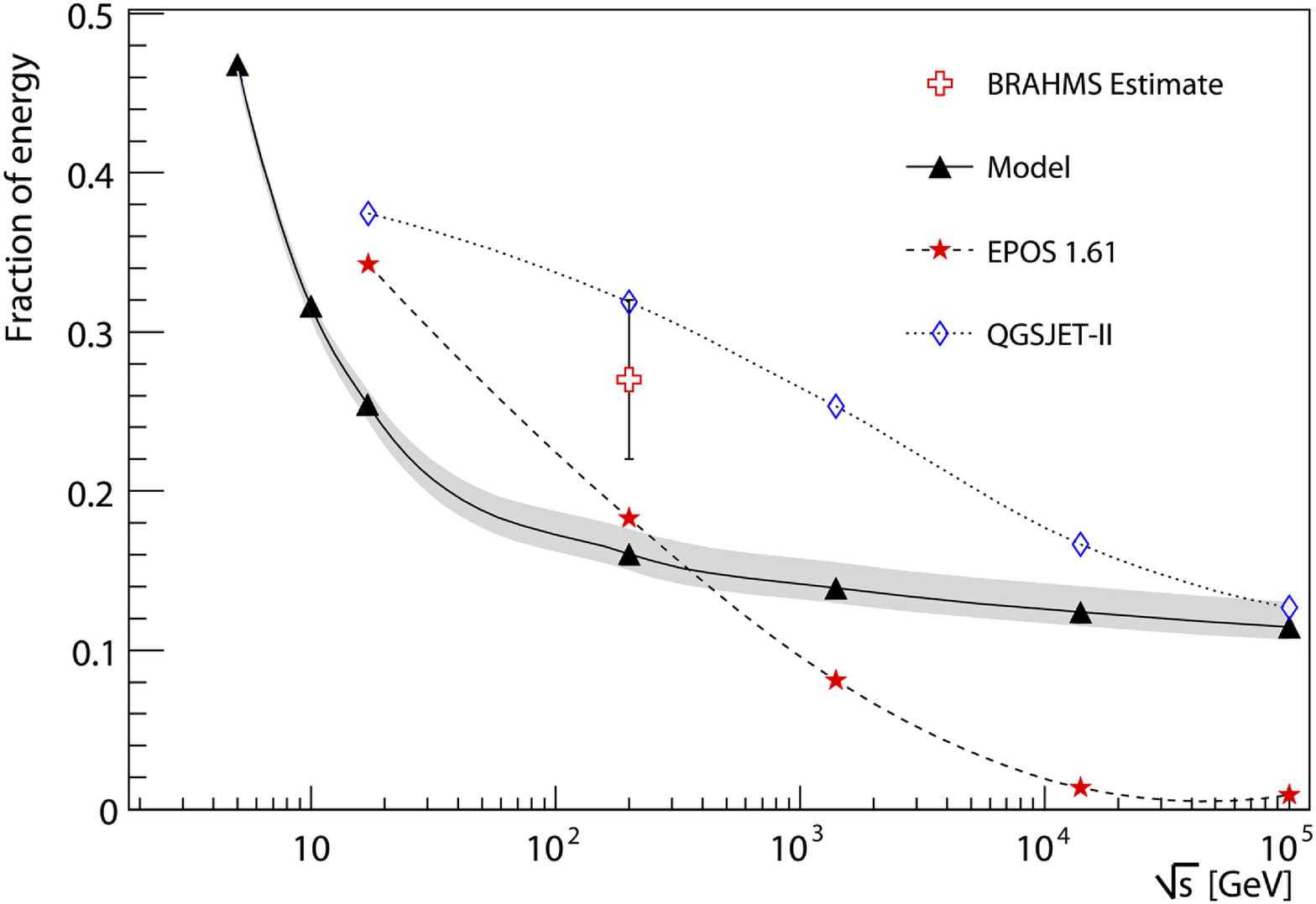, angle=0, width=0.4\textwidth}
\caption{Evolution of the fraction of energy carried by the Net-Baryon with $\sqrt{s}$.
The prediction of the present model (with a shaded band corresponding to the 1$\sigma$ variation
of the fit parameters) is shown, together with the results obtained with EPOS and QGSJET-II.
The data point corresponds to the RHIC estimate given in~\cite{Au-centrality}.}
\label{lambda-x}
\end{center}
\end{figure}

Taking the obtained value of $\lambda_v$, 
the predictions of the present model for net-baryon rapidity and
rapidity loss at higher centre-of-mass energies were obtained and are shown in
figure~\ref{predictions}, covering  both the LHC and the high energy
cosmic ray regions. 
In addition, the fraction of the centre-of-mass energy  
carried by the net-baryon as a function of $\sqrt{s}$ 
was computed and is shown in figure~\ref{lambda-x} (together with the 1$\sigma$ bounds). 
The predictions of EPOS 1.61 and QGSJET-II.03 are also shown.
According to~\cite{Au-centrality}, RHIC data point to about 27\% of the
initial energy remaining in the net-baryon after the collision. This result is also
shown in the figure.
It is apparent that EPOS, QGSJET-II and the present model show rather different trends.
The present model and EPOS agree reasonably well at RHIC energies, where they are both slightly 
below the measured value. At these energies, QGSJET-II is somewhat above, but equally compatible
with data.
At higher energies, while in EPOS 1.61 the net-baryon is essentially zero,
in the present model and in QGSJET-II.03 a sizable amount of energy is still associated to the net-baryon.
It should be noted that high energy effects such as string percolation may 
change these predictions~\cite{percol}.

\section{Summary and conclusions}
\label{sec:conclusions}

A simple but consistent model for net-baryon production in high energy hadron-hadron, 
hadron-nucleus and nucleus-nucleus collisions 
was presented. The basic ingredients of the model are valence string formation based on standard PDFs 
with QCD evolution and string fragmentation via the Schwinger mechanism.
The results of the model were presented and compared with data from RHIC, SPS and AGS, 
and with existing models, namely QGSJET-II and EPOS 1.61. 
The obtained results show that a good 
description of the main features of net-baryon data is achieved on the basis of this 
simple model, in which the fundamental production mechanisms appear in a transparent way.

The free parameters in the model, the effective $Q^2$ and the number of participating
nucleons, were fitted to net-baryon data at $\sqrt{s} \simeq $~5, 17 and 200 GeV. A good fit to
data is obtained within this model, and the values obtained for the number of 
participants are in agreement with what we would expect from the literature.

Using the (scarce) net-baryon data at different centre-of-mass energies, a relation
between the effective momentum transfer and the centre-of-mass energy was motivated 
and a prediction was obtained and extrapolated to higher energies for the evolution
with $\sqrt{s}$ of the fraction of the initial energy carried away by the net-baryon.
A sizable amount of energy may be associated to the net-baryon,
even at high energies.

\vspace{3 mm}

\section*{Acknowledgments}

We thank K. Werner for kindly providing us the EPOS 1.61 code and for 
useful discussions and advice on its usage and results. 
We thank S. Ostapchenko for kindly providing us a QGSJET-II.03
standalone heavy-ion version and for useful discussions.
We thank N. Armesto for useful discussions.
J. A-M is supported by the ``Ram\'on y Cajal'' program, and also by
Ministerio de Educaci\'on y Ciencia (2004-01198), CICYT (FPA 2005-01963),
Xunta de Galicia (2003 PX043, 2005 PXIC20604PN), and Feder Funds, Spain.
R. Concei\c{c}\~ao and J. G. Milhano acknowledge the support of 
FCT, Funda\c{c}\~ao para a Ci\^encia e a Tecnologia and Feder Funds, Portugal.

\end{document}